\documentclass[8pt]{article}  \usepackage{times}
\usepackage{graphicx}

\usepackage{color}

\usepackage[round,numbers,sort&compress]{natbib} 

\newcommand{\be}{\begin{equation}}
\newcommand{\ee}{\end{equation}}

\let\baraccent=\= 
\renewcommand{\=}[1]{\stackrel{#1}{=}} 

\begin{document}

\twocolumn[{\LARGE \textbf{Lipid ion channels and the role of proteins\\*[0.2cm]}}
{\large Lars D. Mosgaard and Thomas Heimburg$^{\ast}$\\*[0.1cm]
{\small $^1$Niels Bohr Institute, University of Copenhagen, Blegdamsvej 17, 2100 Copenhagen \O, Denmark}\\

{\normalsize ABSTRACT\hspace{0.5cm} Synthetic lipid membranes in the absence of proteins can display quantized conduction events for ions that are virtually indistinguishable from those of protein channel. By indistinguishable we mean that one cannot decide based on the current trace alone whether conductance events originate from a membrane which does or does not contain channel proteins.  Additional evidence is required to distinguish between the two cases, and it is not always certain that such evidence can be provided. The phenomenological similarities are striking and span a wide range of phenomena: The typical conductances are of equal order and both lifetime distributions and current histograms are similar. One finds conduction bursts, flickering, and multistep-conductance. Lipid channels can be gated by voltage, and can be blocked by drugs. They respond to changes in lateral membrane tension and temperature. Thus, they behave like voltage-gated, temperature-gated and mechano-sensitive protein channels, or like receptors. 

Lipid channels are remarkably under-appreciated. However, the similarity between lipid and protein channels poses an eminent problem for the interpretation of protein channel data. For instance, the Hodgkin-Huxley theory for nerve pulse conduction requires a selective mechanism for the conduction of sodium and potassium ions. To this end, the lipid membrane must act both as a capacitor and as an insulator. Non-selective ion conductance by mechanisms other than the gated protein-channels challenges the proposed mechanism for pulse propagation. Nevertheless, the properties of the lipid membrane surrounding the proteins hardly ever enter into the textbook discussion of membrane models.

Some important questions arise: Are lipid and protein channels similar due a common mechanism, or are these similarities  fortuitous? Is it possible that both phenomena are different aspects of the same phenomenon? Are lipid and protein channels different at all? 

In this review we will document some of the experimental and theoretical findings that show the similarity between lipid and protein channels. We discuss important cases where protein channel function is strongly correlated to lipid properties. Based on some statistical thermodynamics simulations we discuss how such a correlations could come about. We suggest that proteins can in principle act as catalysts for lipid channel formation and that some apparently mysterious correlations between protein and lipid membrane function can be understood in this manner.\\*[0.0cm] }}
]

\noindent\footnotesize {$^{\ast}$corresponding author, theimbu@nbi.dk}\\


\normalsize
\section{Introduction}

The observation of channel-like conduction events in pure lipid membranes is not new but has not received appropriate attention. Yafuso et al. described them in oxidized cholesterol membranes as early as 1974. Further evidence was provided by Antonov and collaborators and  by Kaufmann and Silman in the 1980s. It was shown that channel formation is influenced by temperature and pH. Goegelein and Koepsell showed in 1984 that one can block such lipid channels with calcium, and Blicher et al. showed that one can block lipid membrane channels with general anesthetics. They may be mildly selective for ions following the Hofmeister series (the above is reviewed in Heimburg, 2010 \cite{Heimburg2010}). Further, lipid channels can be gated by voltage \cite{Blicher2012}. In the last decade, Colombini and collaborators described channel-like events in membranes containing ceramides and sphingolipids \cite{Siskind2002}. 

Synthetic lipid membranes can display melting transitions of their chains \cite{Heimburg2007a}. The transition is accompanied by an absorption of heat, and a change in entropy due to the disordering of the lipid chains. Biological membranes possess such melting transitions, too, typically at temperatures about 10 degrees below physiological temperature \cite{Heimburg2007a}. Chain melting has been described for several bacterial membranes but also for lung surfactant and nerve membranes. Electrophysiological experiments on synthetic membrane suggest that the spontaneous appearance of pores in membranes is related to thermal fluctuations, which are known to approach a maximum in the transition regime. The likelihood of pore opening and the respective open lifetimes are characterized well by the well-known fluctuation-dissipation theorem (FDT)\cite{Kubo1966} and its applications to membrane pores \cite{Heimburg2010}. In essence, the FDT establishes connections between the fluctuations of extensive variables such as enthalpy, volume and area, and the conjugated susceptibilities: heat capacity, isothermal volume and area compressibility, respectively. For instance, wherever the heat capacity is high, the fluctuations in enthalpy are large. Similarly, large volume and area fluctuations imply a high volume or lateral compressibility. Volume and area fluctuations are strongly coupled to the enthalpy fluctuations \cite{Heimburg2007a}. Therefore, the membrane becomes highly compressibly (i.e., soft) close to transition, and the formation of defects is facilitated. For this reason, close to transitions one expects the spontaneous formation of pores in the lipid membrane, and a significant increase of the permeability for ions and small molecules. Such changes in permeability close to transitions were in fact observed experimentally\cite{Heimburg2010}. The presence of melting transitions in biological membranes therefore makes it highly likely that lipid ion channels are also present in living cells. An important implication of the FDT is that fluctuations are directly coupled to fluctuation lifetimes \cite{Seeger2007}, which implies that the open lifetime of lipid membrane pores is maximum close to the transition. 

Changes in experimental conditions can shift the melting point. For instance, drugs such as the insecticide lindane or the anesthetic octanol lower transition temperatures and therefore influence the permeability of membranes \cite{Heimburg2010}. Due to their effect on the physics of membranes, such drugs can `block' lipid channels without binding to any particular receptor. Similarly, hydrostatic pressure shifts transitions upwards \cite{Heimburg2007a}.  

It seems to be increasingly accepted that the composition of lipid membranes can influence the channel activity of proteins \cite{Schmidt2006}. Interestingly, critical phenomena such as the ones described above for synthetic membranes have also been found for protein channels embedded in synthetic lipid membranes. The characteristics of some of these channels are highly correlated with the chain melting transition in the surrounding membrane. For instance, the KcsA potassium channel when reconstituted into a synthetic membrane displays a mean conductance that exactly reflects the heat capacity profile of the membrane \cite{Seeger2010}. Simultaneously, the open lifetime of the channel is maximum in the membrane transition. Thus, the properties of membranes containing this channel protein accurately reflect the physics of the fluctuations in the lipid membrane. Very similar phenomena were found for the sarcoplasmic reticulum calcium channel \cite{Cannon2003}. 
\begin{figure}[!t]
	\centering
		\includegraphics[width= 1 \linewidth]{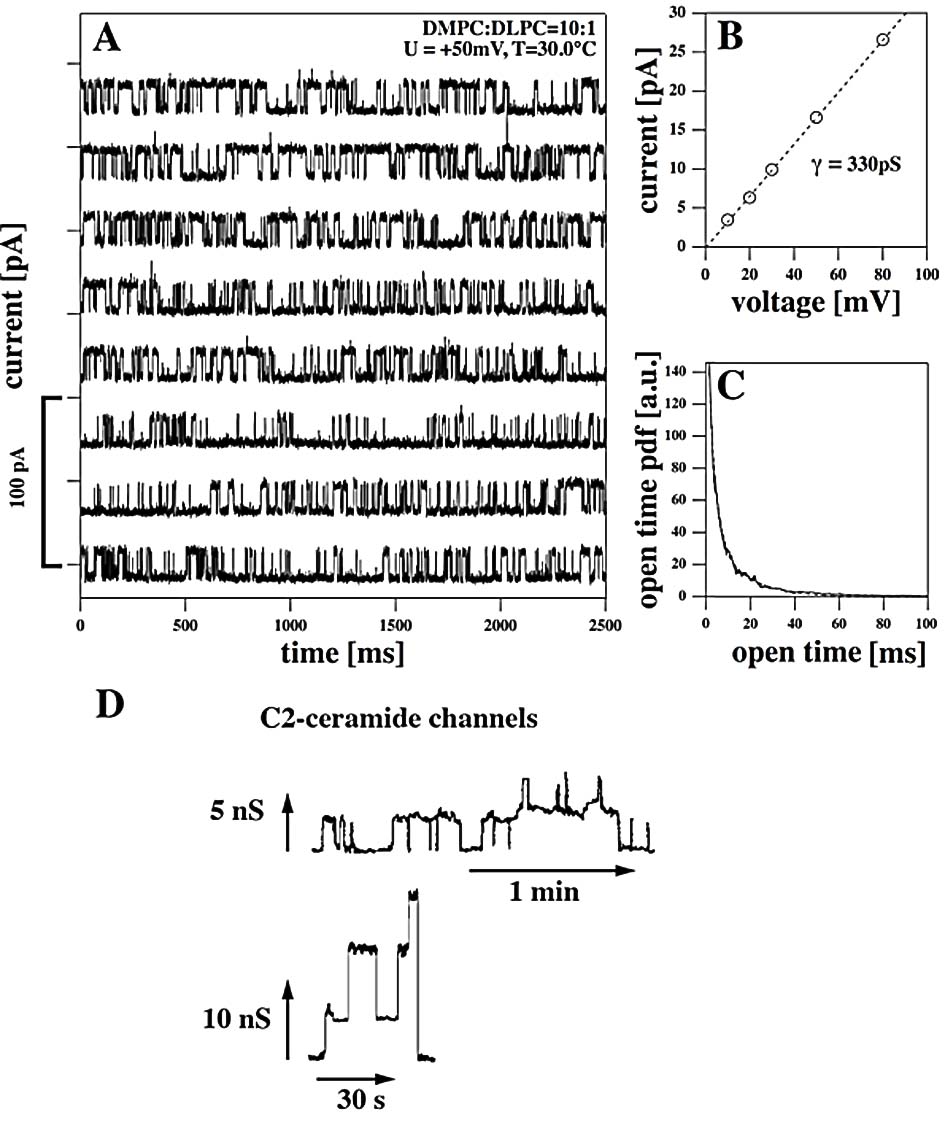}
	\parbox[c]{8cm}{ \caption{\textit{Lipid ion channels in a synthetic membrane (DMPC:DLPC=10:1 in 150mM KCl, 30$^\circ$C). \textbf{A. }Characteristic current trace recorded at 50mV. It was stable for more than 30 minutes. \textbf{B. }The single-channel  I-V profile is linear, resulting in a conductance of 330 pS. \textbf{C. }Lifetime distribution of the channels. Adapted from Laub et al, 2012 \cite{Laub2012}. \textbf{D.} Ceramide channels in soybean lipids (asolectin) in 1M KCl. From Siskind et al., 2002 and 2003 \cite{Siskind2002, Siskind2003}. 
	}
	\label{Figure_02}}}
\end{figure}
\section{Lipid ion channels}
\subsection{Single channels in protein-free membranes}
Fig. \ref{Figure_02}A shows a typical channel event in a synthetic lipid membrane (DMPC: DLPC=10:1, 150mM NaCl) recorded around 30$^\circ$C \cite{Laub2012}. These channels display a conductance of about 330 pS and lifetimes on the order of 1-100ms. Such conductances and lifetimes are not untypical for protein channels, too. In fact, it would be difficult to distinguish the traces in Fig. \ref{Figure_02} from protein channels in the absence of independent information available. Further examples were reviewed in Heimburg, 2010 \cite{Heimburg2010}. 

Fig. \ref{Figure_02}D shows recordings from so-called ceramide channels. Such channels have been investigated in much detail by the Colombini group from the University of Maryland. Channel events in protein-free membranes containing a small amount of ceramide lipids look similar to the lipid pores in Fig. \ref{Figure_02}. Ceramide channels seem to be distinct from other lipid membrane pores since they display much larger conductances and much longer lifetimes than the data in Fig. \ref{Figure_02} (A-C). 

\subsection{Comparison of lipid and TRP protein ion channels}

The literature contains copious examples for protein channel conductance. As mentioned, many of these data are similar in appearance to the lipid channels. This is demonstrated in the following for TRP channels that were over-expressed in human embryonic kidney (HEK) cells \cite{Laub2012}. 
\begin{figure}[!hb]
	\centering
		\includegraphics[width= 1 \linewidth]{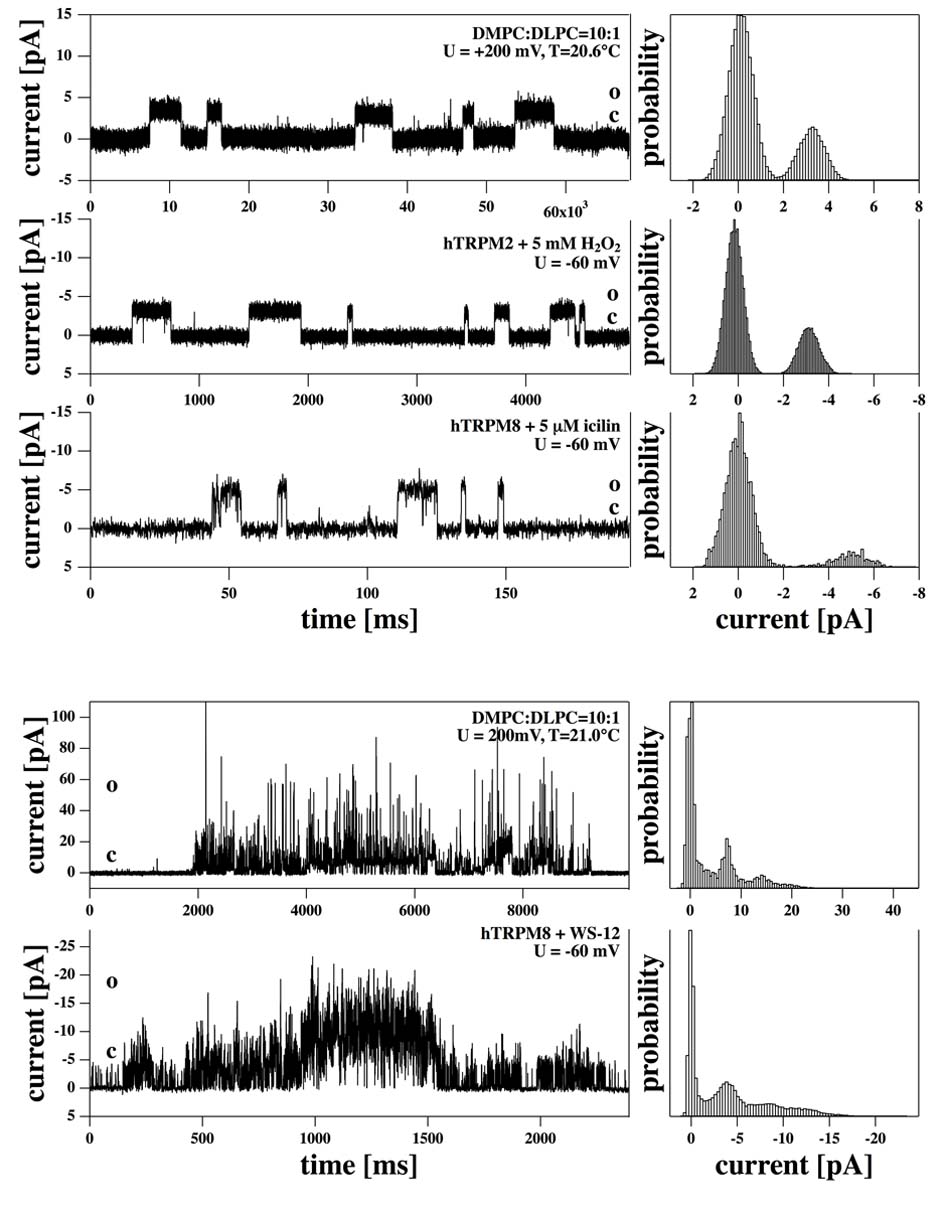}
	\parbox[c]{8cm}{ \caption{\textit{\textbf{Top: }The comparison of channel events from synthetic lipid membranes (top) and from protein-containing cell membranes (center: TRPM2, bottom:TRPM8) demonstrate the phenomenological similarity of both stepwise conductance and probability distribution of current events \cite{Laub2012}. \textbf{Bottom: }Conduction burst in a synthetic membrane (top) and in a HEK cell membrane containing the TRPM8 channel (bottom) \cite{Laub2012}. The current histograms are nearly identical.}
	\label{Figure_02b}}}
\end{figure}
Fig. \ref{Figure_02b} (top) shows short but representative time-segments of channel recordings from a synthetic lipid preparation (top trace), HEK cells containing the TRPM2 (center trace) and TRPM8 (bottom trace) channels (data from Laub et al. \cite{Laub2012}). The order of magnitude of the conductance and the current histograms are very similar. Equally similar traces were found for conduction bursts, flickering traces and other phenomena typical for protein conductance \cite{Laub2012}.  Fig. \ref{Figure_02b} (bottom) a conduction burst in a synthetic membrane (top) is compared to a burst of the TRPM8 channel activity. The current histogram of the two bursts is very similar, both in absolute currents and in the peak areas. It is again difficult to distinguish the data from the synthetic and cell membranes without independent experimental evidence. It is likely that this statement is generally true.

\subsection{Pore geometries}
\begin{figure}[!htb]
	\centering
		\includegraphics[width= 1 \linewidth]{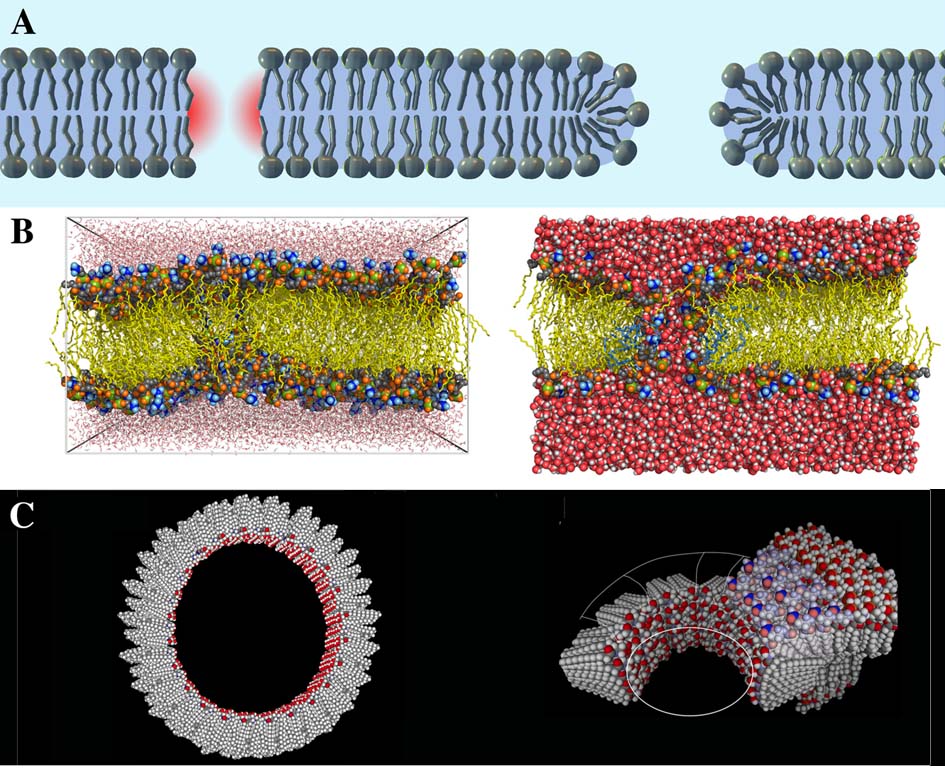}
	\parbox[c]{8cm}{ \caption{\textit{Hydrophobic and hydrophilic pores in lipid membranes. \textbf{A. }Schematic drawing with hydrophobic pore on the left and hydrophilic pore on the right. \textbf{B. }Molecular dynamics simulation of a membrane subject to a voltage difference of 2 V\cite{Boeckmann2008}.  Here, the the hydrophobic pore (left) consists of a file of water molecules spanning through the membrane. Such pores are thought to be dynamic structures caused by fluctuations in the lipid bilayer. \textbf{C. }Hypothetical structure of a ceramide channel.\cite{Samanta2011}.}
	\label{Figure_05}}}
\end{figure}

Lipid pores are transient in time and probably due to area fluctuations in the membrane. There exists no experimental evidence for a well-defined pore geometry and size from electron microscopy or other nanoscopic techniques. Glaser and collaborators \cite{Glaser1988} proposed two kinds of pores: The hydrophobic pore shown in Fig. \ref{Figure_05}A (left) and the hydrophilic pore shown in Fig. \ref{Figure_05}A (right). The hydrophobic pore is a area density fluctuation without major rearrangement of its lipids. Water in the pore is in contact with hydrophobic hydrocarbon chains.  The hydrophilic pore (Fig. \ref{Figure_05}A, right) is linked to a rearrangement of the lipids so that contact with water is avoided. Such pores are thought to be more stable and long-lived. Based on elastic considerations, the pore diameter was estimated to be of order 1nm \cite{Glaser1988}, suggesting a well-defined channel conductance. Both, hydrophilic and hydrophobic pores have been found in MD simulations \cite{Boeckmann2008} (Fig. \ref{Figure_05}B). Application of voltage across the membrane initially forms a hydrophobic pore which subsequently develops into a hydrophilic pore.  Due to the lack of direct observation, these geometries must be regarded as tentative.
The proposed structure of ceramide channels consists rather of a stable geometry with well-defined molecular order. Considering their longer lifetimes and larger conductance, it is not clear whether these events are comparable to the lipid ion channel events described in Fig. \ref{Figure_02}.

\section{Transitions in the membrane and permeability maxima}
\begin{figure*}[t]
	\centering
		\includegraphics[width= 0.9 \linewidth]{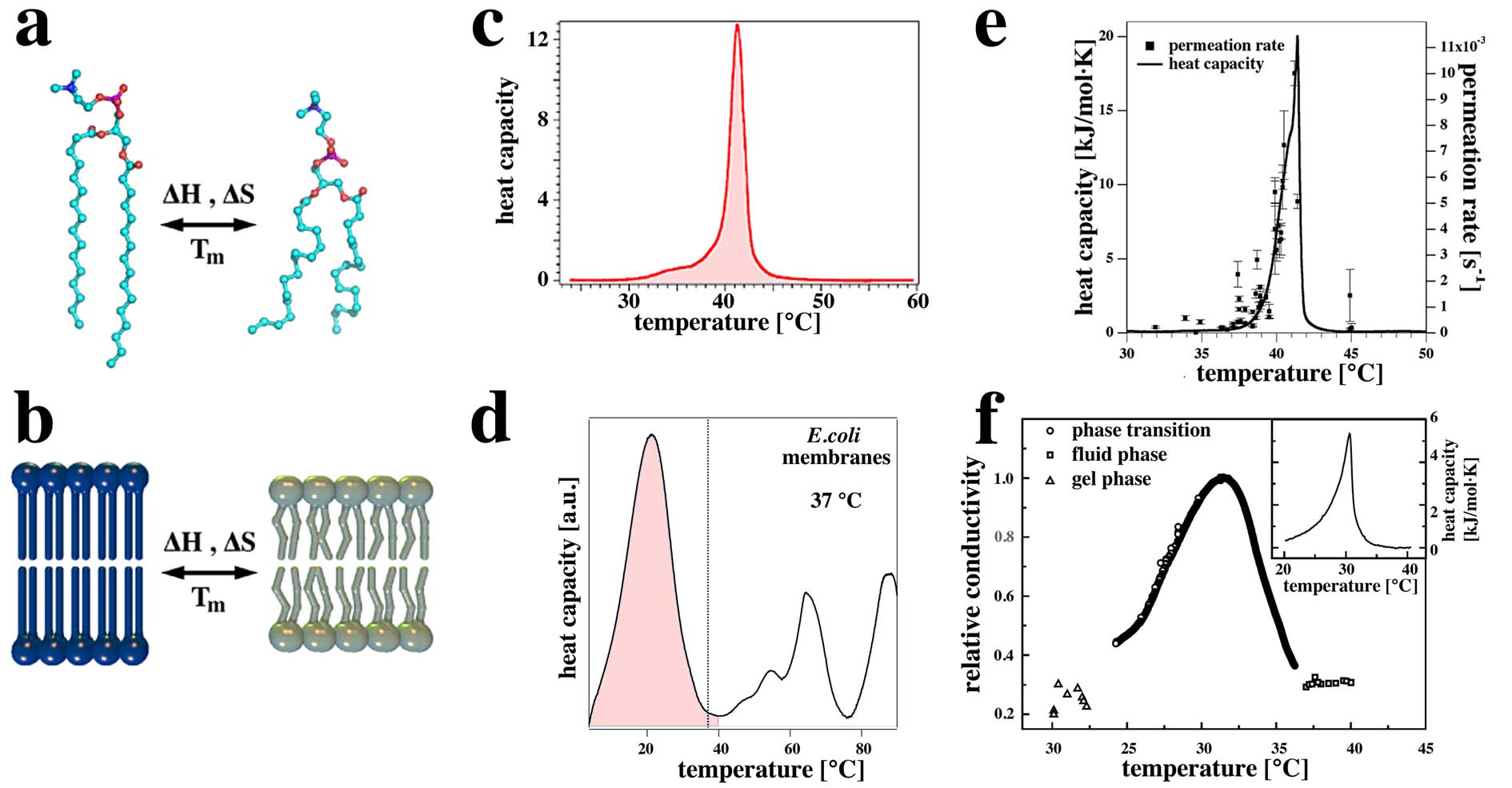}
	\parbox[c]{16cm}{ \caption{\textit{Melting of lipid membranes and permeability changes. \textbf{a, b.} Schematic representation of the melting process in a lipid membrane \cite{Heimburg2007a}. \textbf{c. }Calorimetric melting profile of a dipalmitoyl phosphatidylcholine (DPPC) membrane. \textbf{d. }Melting profile of native E. \textit{coli} membranes \cite{Heimburg2007a}. The peak shaded in red is the lipid melting peak, situated about 10-15 degrees below growth temperature (dashed line). \textbf{e.} Permeability of a synthetic lipid membrane for a fluorescence dye compared to its heat capacity \cite{Blicher2009}. \textbf{f.} Conductance of a synthetic lipid membrane for ions compared to the heat capacity \cite{Wunderlich2009}}.
	\label{Figure_06}}}
\end{figure*}

Lipid membranes display melting transitions. In these transitions the membranes remain intact, but enthalpy $\Delta H$ and entropy $\Delta S$ change at a defined temperature $T_m=\Delta H/\Delta S$. This is schematically displayed in Fig. \ref{Figure_06} (a, b). One can investigate such transitions in calorimetry by recoding the heat capacity, $c_p = (dH/dT)_p$. The heat capacity profile shown in panel c is from synthetic vesicles of dipalmitoyl phosphatidylcholine (the lipid in panel a) with a melting temperature of $\sim$41$^\circ$C. Panel d (shaded area) shows the $c_p$-profile of native E. \textit{coli} membranes with a lipid melting peak close to physiological temperatures ($T_m\approx 22^\circ$C). This has also been reported for other bacterial membranes, for lung surfactant \cite{Heimburg2007a}, and for nerve membranes from rat brains, where the heat capacity maximum is always found in the range from 20-30$^\circ$C. It seems likely that chain melting close to physiological conditions is a generic property of cells.
The integrated heat capacity yields both $\Delta H$ and the $\Delta S$. Simultaneously, the volume of the membrane changes by about 4\% and the area by about 24\% \cite{Heimburg1998}. The fluctuation-dissipation theorem (FDT) states that heat capacity $c_p$, volume and area compressibility 
$\kappa_T^V$ and $\kappa_T^A$ are given by
\begin{eqnarray}
\label{eq:T01}
c_p=\frac{\left<H^2\right>-\left<H\right>^2}{RT^2} \quad;\quad\nonumber\\ 
\kappa_T^V=\frac{\left<V^2\right>-\left<V\right>^2}{\left<V\right>RT} \quad;\quad \\
\kappa_T^A=\frac{\left<A^2\right>-\left<A\right>^2}{\left<A\right>RT }\quad,\quad\nonumber
\end{eqnarray}
i.e., they are related to enthalpy, volume and area fluctuations.

Empirically, it was found that
\begin{equation}
\label{eq:T03}
\kappa_T^V=\gamma_V \cdot c_p\quad;\quad \kappa_T^A=\gamma_A \cdot c_p \;,
\end{equation}
where $\gamma_V\approx 7.8\cdot 10^{-10}$m$^2/N$ and $\gamma_A\approx 0.89$ m/N are material constants \cite{Heimburg1998, Ebel2001, Pedersen2010}. As a consequence, membranes become very soft in transitions. The application of the FDT to lipid membranes is discussed in detail by Heimburg (2010) \cite{Heimburg2010}.

Several previous studies reported that membranes become more permeable in the transition regime \cite{Heimburg2010}. In order to create a pore, work $\Delta W(a)$ has to be performed: 
\begin{equation}
\Delta W(a)=\frac{1}{2\kappa_T^A A_0}  a^2 \;,
\label{eq:T04}
\end{equation}
where $a$ is the area of the pore and $A_0$ is the area of the overall membrane. Here, the line tension of the pore circumference\cite{Winterhalter1987, Glaser1988} is not explicitly contained for reasons discussed in Blicher et al. \cite{Blicher2009}. Since $\kappa_T^A$ has a maximum in  the transition, the likelihood of finding a pore is enhanced and the permeability is high. This is shown for fluorescence dyes and ions in Fig. \ref{Figure_06} (e and f). 

\subsection{Relaxation timescales and lipid channel lifetimes}
The lifetime of fluctuations is also described by the fluctuation-dissipation theorem \cite{Kubo1966}. Generally, larger fluctuations are associated with larger timescales. 
\begin{figure}[!bht]
	\centering
		\includegraphics[width= 1 \linewidth]{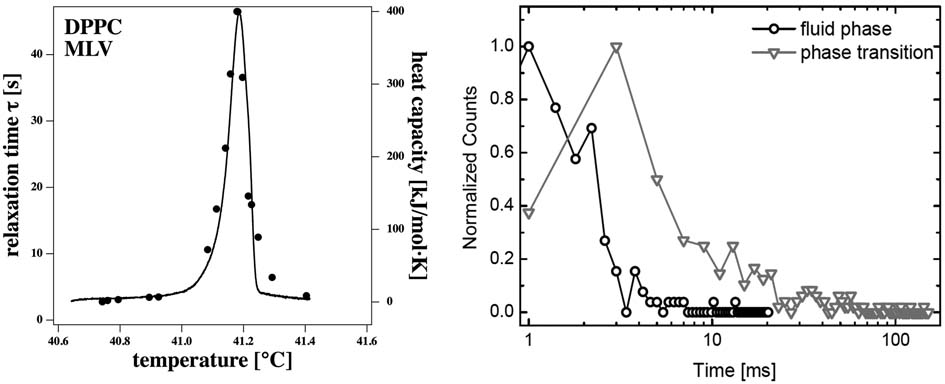}
	\parbox[c]{8cm}{ \caption{\textit{\textbf{Left: }Relaxation time scale of DPPC MLV compared to the heat capacity \cite{Seeger2007}. \textbf{Right: }Open lifetimes of channel events in a D15PC: DOPC = 95:5 mixture in the fluid phase and in the phase transition \cite{Wunderlich2009}.}
	\label{Figure_15}}}
\end{figure}
In the case of single lipid membranes close to transitions it has been shown that relaxation times are reasonably well described by an single-exponential process with a time constant of
\begin{equation}
\label{eq:LT_05}
\tau=\frac{T^2}{L}\Delta c_p \;,
\end{equation}
where $L$ is a phenomenological coefficient  with $L=6.6\cdot 10^{8}$ J K/mol s for multilamellar lipid vesicles (MLV) of synthetic lipids \cite{Seeger2007}. In nonequilibrium thermodynamics it is assumed that the relaxation and the fluctuation timescales are identical. Therefore, the time scale $\tau$ is intimately related to the mean open time of lipid channels. Fig. \ref{Figure_15}\,(left) shows that DPPC MLV display a $\Delta c_p$ maximum of about 400 kJ/mol K, leading to a relaxation time $\tau$ of 45 seconds. The half width of this transition is of the order of 0.05-0.1 K.  

The half width of the heat capacity profile  of lung surfactant is much broader (order 10 K). The heat capacity at maximum is about 300 times smaller than for DPPC, and one expects a maximum relaxation time on the order of 100\,ms. Similar numbers are expected for the transitions in E. coli and other bacterial membranes, and the membranes of nerves. Interestingly, 1 - 100\,ms is the timescale of the open lifetime of protein channels -- as well as that of lipid channels. The right hand panel of Fig. \ref{Figure_15} shows the distribution of lipid channel lifetimes within and above the melting range of a synthetic lipid mixture \cite{Wunderlich2009}. It can be seen that the lifetime increases 5 to 10-fold in the transition range, which is consistent with the above considerations.  Below, we will show that the KcsA potassium channel displays similar dependence on the melting transition in the surrounding membrane. 

\subsection{Gating of lipid channels}
The likelihood of finding lipid pores depends on experimental conditions due to the dependence of the melting point on all intensive thermodynamic variables.

The permeability must be considered as primarily due to pore formation. Therefore, the likelihood of finding channels is correlated with changes in the intensive variables. In analogy with the nomenclature of protein channels, we will call this effect `gating'. Gating implies that the open probability of lipid channels depends in a very general sense on the intensive thermodynamic variables. It has been shown experimentally that lipid channels can be gated by \cite{Heimburg2010} \begin{itemize}
  \item temperature   (temperature-sensing).
  \item lateral pressure or tension  (mechanosensitive-gating).
  \item general anesthetics  (gating by drugs). 
  \item calcium and pH , i.e., chemical potential differences of calcium and protons.
  \item voltage (voltage-gating), discussed below.
\end{itemize} 
Note that these variables have also been reported to control protein channels, e.g., the temperature sensitive TRP channels \cite{Voets2004}, mechanosensitive channels \cite{Suchyna2004}, the effect of general anesthetics on the nicotinic acetyl choline receptor \cite{Bradley1984}, calcium channels \cite{Cannon2003}, the pH-dependent \cite{Seeger2010} and voltage-gated \cite{Schmidt2006} KcsA potassium channels.

\subsection{The effect of voltage}\label{voltage}
\begin{figure}[!htb]
	\centering
		\includegraphics[width= 1 \linewidth]{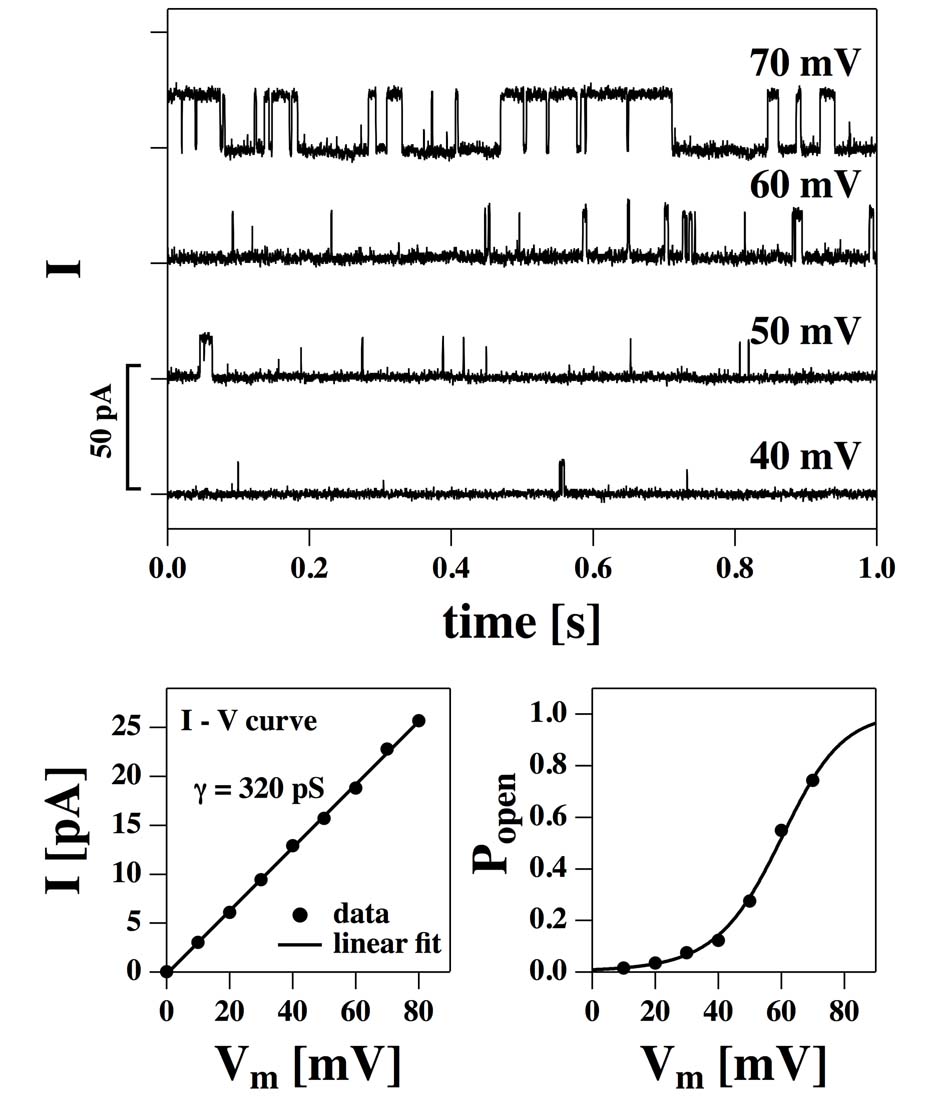}
	\parbox[c]{8cm}{ \caption{\textit{Voltage-gating in a DMPC:DLPC=10:1 membrane at 30$^\circ$C in 150mM KCl.  \textbf{Top: }Current-traces  at four voltages showing an increase of single-channel conductance with voltage and an increased likelihood of channel formation. \textbf{Bottom, left: }The corresponding linear single-channel current-voltage relation indicating a single-channel conductance of $\gamma$=320 pS. \textbf{Bottom, right: }Open probability as a function of voltage.}
	\label{Figure_08}}}
\end{figure}

At suitable voltages, one can induce single channel events in the synthetic membrane (Fig. \ref{Figure_08}). In contrast to the conductance of the overall membrane, the single channel conductance is constant and leads to a linear current-voltage relation. The open probability increases as a function of voltage.

The phase diagram of a membrane as a function of voltage was recently given by Heimburg \cite{Heimburg2012}. If one regards the membrane as a capacitor, one can calculate the force on the membrane due to electrostatic attraction. This force can reduce the thickness of the membrane.  It can thus change the melting temperature and potentially create holes above a threshold voltage \cite{Crowley1973, Winterhalter1987}. The electrostatic force, $\mathcal{F}$, exerted by voltage on a planar membrane is given by 
\begin{equation}
\label{eq:01}
\mathcal{F}=\frac{1}{2}\frac{C_m V_m^2}{D} \;
\end{equation}
where $C_m$ is the membrane capacitance, $V_m$ is the transmembrane voltage and $D$ is the membrane thickness \cite{Heimburg2012}. This force reduces the thickness of the membrane \cite{Blicher2012}. The electrical work performed on the membrane by a change in thickness from $D_1$ to $D_2$ is
\begin{equation}
\label{eq:02}
\Delta W_{el}=\int_{D_1}^{D_2}\mathcal{F}dD \propto V_m^2 \;.
\end{equation}
\begin{figure}[!b]
	\centering
		\includegraphics[width= 1 \linewidth]{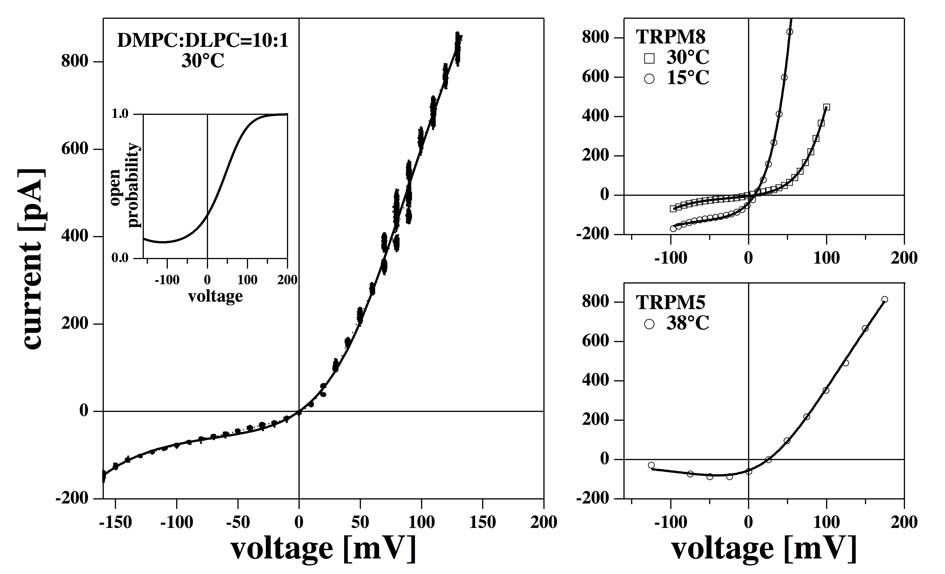}
	\parbox[c]{8cm}{ \caption{\textit{Current voltage-relations. \textbf{Left: }A synthetic membrane (DMPC:DLPC=10:1). The insert is the open probability of a membrane pore. The solid line represents a fit to eq. \ref{eq:05} with $E_0=0 $\,V. \textbf{Right: }I-V profiles of TRPM8 channels in HEK cells, adapted from Voets et al. 2004 \cite{Voets2004} (top panel) and of TRPM5 in HEK cells, adapted from Talavera et al. 2005 \cite{Talavera2005}. The solid lines are fits to eq. \ref{eq:05}.  }
	\label{Figure_07}}}
\end{figure}

It can thus be assumed that the free energy of pore formation, $\Delta G$, is related to the square of the voltage and to the elastic constants of the membrane \cite{Blicher2012} with 
\begin{equation}
\label{eq:03}
\Delta G=\Delta G_0 +\alpha V_m^2 \;,
\end{equation}
where $\Delta G_0$ is the free energy difference between open and closed pores in the absence of voltage and $\alpha$ is a constant. $\Delta G_0$ reflects the elastic properties of the membrane that depend on composition, temperature and pressure. For asymmetric membranes one obtains
\begin{equation}
\label{eq:03b}
\Delta G=\Delta G_0 +\alpha (V_m-V_0)^2 \;,
\end{equation}
where the offset voltage $V_0$ is due to membrane curvature or to a different lipid composition in the two membrane leaflets \cite{Alvarez1978}. 
The probability, $P_{open} (V_m)$, of finding an open pore in the membrane at a fixed voltage is given by
\begin{equation}
\label{eq:04}
P_{open}(V_m)=\frac{K(V_m)}{1+K(V_m)}\;;\; K(V_m)=\exp\left(-\frac{\Delta G}{kT}\right) \;,
\end{equation}
where $K(V_m)$ is the voltage-dependent equilibrium constant between open and closed states of a single pore. 

The current-voltage relation for the lipid membrane is proportional to the likelihood of finding an open channel for a given voltage:  
\begin{equation}
\label{eq:05}
I_m=\gamma_p\cdot P_{open}\cdot (V_m-E_0)\;,
\end{equation}
where $\gamma_p$ is the conductance of a single pore and $E_0=$ is the Nernst potential.
While the voltage $V_0$ reflects the asymmetry of the membrane, $E_0$ reflects the asymmetry of the ion concentrations of the buffer solution. If the aqueous buffer is the same on both sides of the membrane, the Nernst potential is zero.  

Fig. \ref{Figure_07} (left) shows the I-V profile of a lipid membrane made of  a mixture of DMPC and DLPC \cite{Laub2012} and a fit given by the formalism given by eqs. \ref{eq:03b} to \ref{eq:05} ($V_0=-110$mV, $\gamma_m=6.62$ nS, $\Delta G_0=5.2$ kJ/mol, and $a=-248$ kJ/mol$ \cdot$V$^2$). This fit reproduces the experimental profile. For comparison, the right hand panel of Fig. \ref{Figure_07} shows the current-voltage relationships of two proteins from the TRP family. Members of this family of ion channels have been reported to respond to environmental stimuli such as temperature, membrane tension, pH and various drugs \cite{Wu2010}. No particular structure for these channels is known, and no very well-defined selectivities for ions have been reported. Fig. \ref{Figure_07} (right panels) show data for the current-voltage relationship of TRPM8 at two temperatures and TRPM5 adapted from publications of B. Nilius' group \cite{Voets2004, Talavera2005}. The I-V profiles look quite similar to that obtained from the synthetic membrane. The solid lines in Fig. \ref{Figure_07} (right panels) are fits to the above formalism using parameters of similar order of magnitude as used for the synthetic membrane. The quality of these fits indicates that the TRP channel conductance is well described as unspecific pore formation in an asymmetric membrane caused by the charging of the membrane capacitor.

\section{The role of proteins}
Here, we consider several cases where proteins and lipid pores display similar dependences on intensive variables.
\subsection{Temperature dependence of lipid pores, temperature sensing protein channels and van't Hoff law}

Sensitivity to temperature is one of the most prominent properties of the TRP channels \cite{Talavera2005}.  Such channels display a temperature sensitivity over a temperature range of 10 K that is similar to the width of melting transitions in many biomembranes. For instance, the TRPM8 channel is activated at temperatures about 10 degrees below body temperature, just where the maximum of the melting profile of many cell membranes is found.  While this may be coincidental, the assumption of temperature-sensing macromolecules is problematic as we discuss below.
\begin{figure}[!htb]
	\centering
		\includegraphics[width= 1 \linewidth]{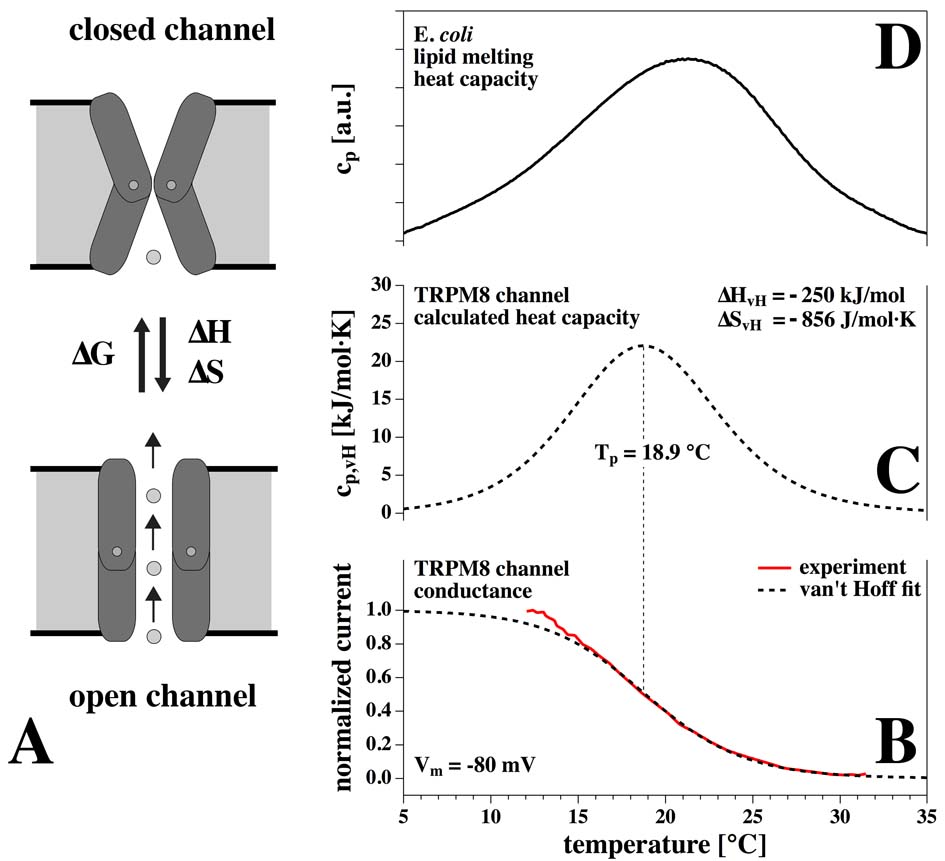}
	\parbox[c]{8cm}{ \caption{\textit{\textbf{A. }Schematic drawing of a possible equilibrium between an open and a closed state of a channel protein. The equilibrium is defined by Gibbs free energy, enthalpy and entropy differences ($\Delta G$, $\Delta H$ and $\Delta S$, respectively). \textbf{B. }Van\ 't Hoff analysis of the temperature dependent conductance of a TRPM8 channel in HEK cells at -80mV (adapted from \cite{Voets2004}). It leads to $\Delta H$ of -250 kJ/mol and $\Delta S$=-859 J/mol$\cdot$ K for the two-state equilibrium. \textbf{C. }From the analysis in panel B one obtains a heat capacity profile of the transition with a midpoint at 18.9 $^\circ$C. \textbf{D. }For comparison, the lipid melting profile of native E. coli membranes is shown. It displays similar transition width and midpoint.}
	\label{Figure_09}}}
\end{figure}

Assume a channel protein with open and closed states as shown in Fig. \ref{Figure_09}A. These states correspond to distinct protein conformations. The equilibrium constant between the two states is $K=\exp{(-\Delta G/RT)}$ with $\Delta G=\Delta H- T\Delta S$.  The likelihood for finding open and 
closed states is  
\begin{equation}
\label{eq:Popen}
P_{open}=\frac{K}{1+K} \qquad\mbox{and}\qquad P_{closed}=\frac{1}{1+K}\, 
\end{equation}
respectively (van\ 't Hoff's law).  On this basis, Talavera et al.\ reported activation enthalpies on the order of 200kJ/mol for TRPM4, TRPM5, TRPM8 and TRPV1 channels \cite{Talavera2005}.  Fig. \ref{Figure_09}B shows a fit of the conductance of a TRPM8 channel \cite{Voets2004} to eq. \ref{eq:Popen}. The corresponding transition enthalpy is $\Delta H=$-250 kJ/mol, and the entropy $\Delta S=$-856 J/mol$\cdot$K.  Fig. \ref{Figure_09}C displays the corresponding heat capacity of the transition in the protein with a maximum at 18.9 $^\circ$C. It is given by the derivative of the fit in Fig. \ref{Figure_09}B multiplied with the van't Hoff enthalpy. 
These numbers are comparable to the total heat of protein unfolding. Typical values are: about 350kJ/mol for staphylococcal ribonuclease, 200 kJ/mol for lysozyme, and 200-500 kJ/mol for metmyoglobin.  However, such transition enthalpies seem highly unlikely for the small conformational change from an open to a closed state.  The conformational change from closed to open state of a protein should have a much smaller enthalpy change and should thus have a much smaller temperature dependence.  Similarly, the claimed difference in the entropy of closed and open states is too large.  According to Boltzmann's equation, $S=k\ln \Omega$ (with $\Omega$ being the degeneracy of states), an entropy difference of -856 J/mol K corresponds to a change of the number of states by a factor of $\Omega=5\cdot 10^{44}$.  A change of this magnitude is plausible for protein denaturation where there is no well-defined unfolded structure but one well defined native conformation.   It is not reasonable for a transition between two states with well-defined function and geometry.  Due to cooperative behavior, however, the melting of the lipid membrane can easily have enthalpies and entropies of the above order. Fig. \ref{Figure_09}D shows the experimental melting profile of E. \textit{coli} membranes. The activation profile of TRPM8 channels is apparently quite consistent with the melting profile of a cell membrane.

\begin{figure}[!htb]
	\centering
		\includegraphics[width= 1 \linewidth]{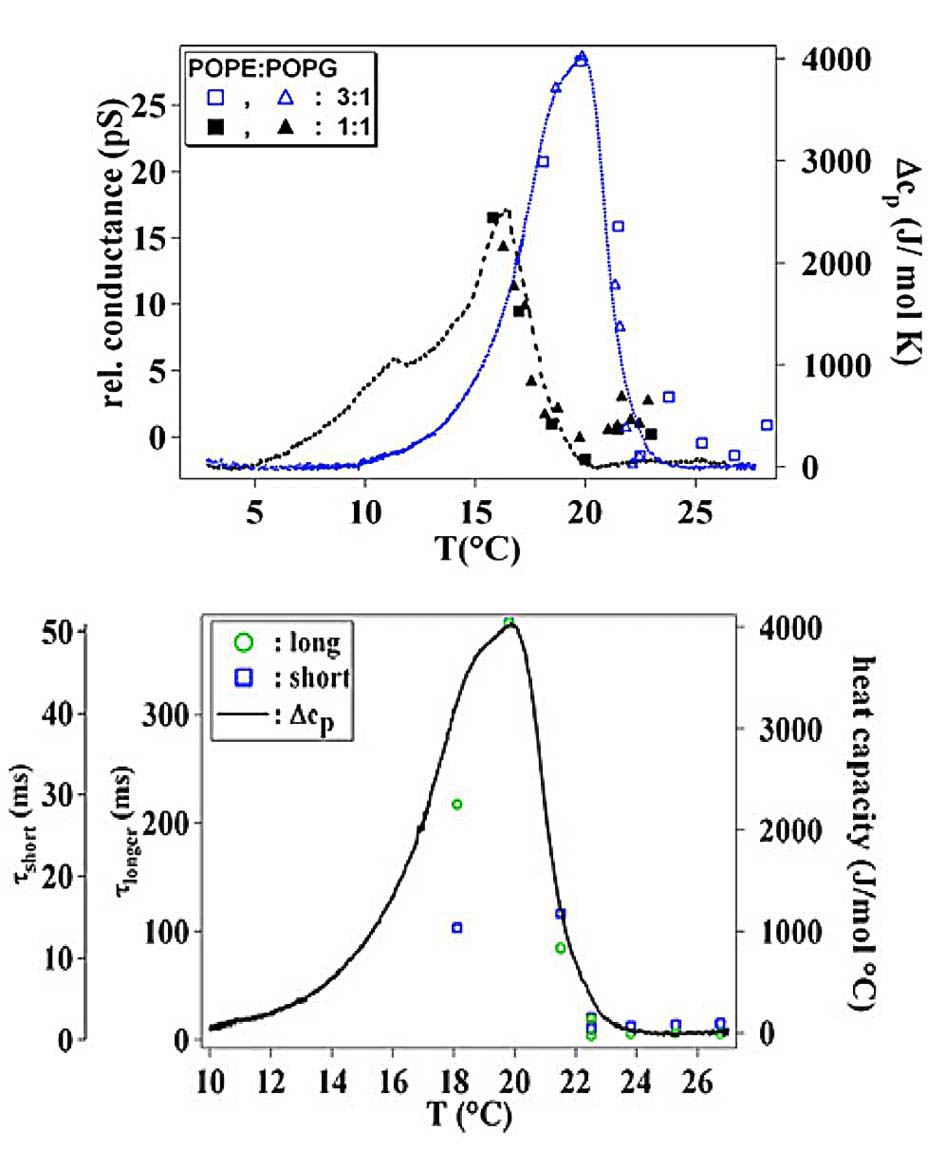}
	\parbox[c]{8cm}{ \caption{\textit{The dependence of KcsA channels on the phase transition of its host membrane. \textbf{Top:} The conductance of the channel in two lipid mixtures \textbf{(symbols)} compared to the respective heat capacity profile of the lipids \textbf{(lines)}. \textbf{Bottom:} Open lifetimes of the KscA-channel \textbf{(symbols)} compared to the heat capacity profile \textbf{(line)}. From Seeger et al. (2010) \cite{Seeger2010}. }
	\label{Figure_12}}}
\end{figure}

\begin{figure*}[!t]
	\centering
		\includegraphics[width= 0.8 \linewidth]{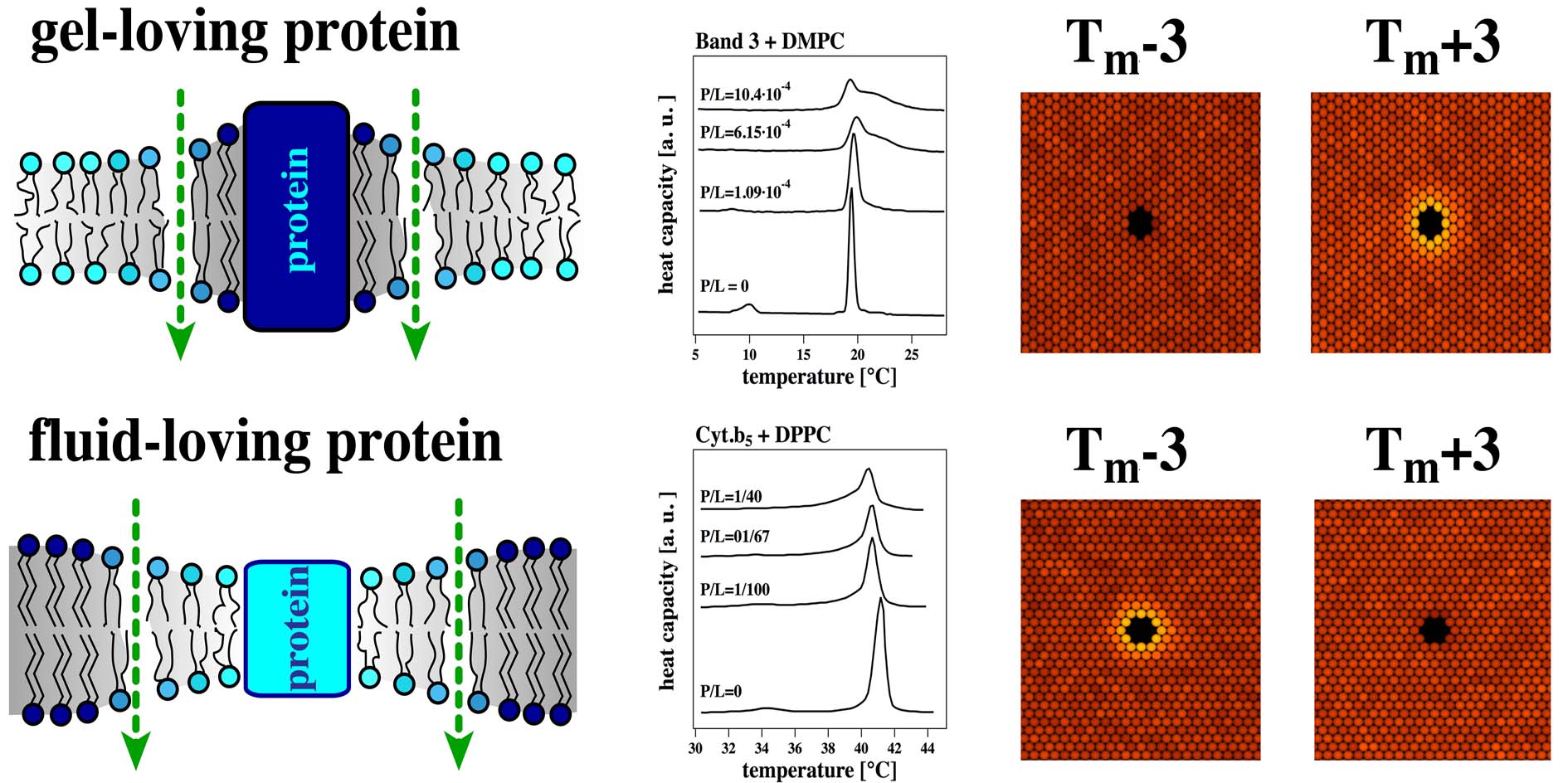}
	\parbox[c]{16cm}{ \caption{\textit{The interaction of proteins with lipid membranes and hydrophobic matching. The top row refers to a protein that favors the solid state, while the bottom row refers to a protein favoring the liquid state.  \textbf{Left: }Schematic drawing of the lipid arrangement around proteins. \textbf{Center: }The influence of proteins on the heat capacity profile of a synthetic membrane (\textbf{Top: }Band 3 protein of erythrocytes in DMPC shifts heat capacity profiles towards higher temperatures. \textbf{Bottom: }Cytochrome B$_5$ in DPPC shifts c$_p$ profiles towards lower temperatures). \textbf{Right: }Monte Carlo simulations of the local fluctuations (yellow indicates larger fluctuations) of the lipid membrane close to the protein. \textbf{Top: }Fluctuations are enhanced at the protein interface above the melting temperature T$_m$ and unaffected below. \textbf{Bottom: }Fluctuations are enhanced below T$_m$ and unaffected above the transition.}
	\label{Figure_14}}}
\end{figure*}

\subsection{Channel proteins and phase transitions in the lipid membrane}
The KcsA potassium channels are pH- and voltage-gated. In order to measure their properties they are frequently reconstituted into synthetic membranes such as POPE:POPG=3:1 mol:mol \cite{Seeger2010}.  It is unclear why this particular lipid mixture is often chosen.  Measurement of the heat capacity reveals that this lipid mixture has a melting profile with a maximum close to room temperature (blue line in Fig. \ref{Figure_12}, top). Interestingly, the measurement of the mean conductance of the KcsA channel reconstituted into this membrane (blue symbols) exactly follows the heat capacity profile. This is not accidental since a change in the lipid composition to POPE:POPG=1:1 reveals that the conductance profile of the channel shifts in the same manner as the heat capacity curve (black symbols). A similar observation can be made for the KcsA channel open times. Typically, the open time distribution is fitted with a biexponential yielding two time constants. These two time constants are plotted in Fig. \ref{Figure_12} (bottom) for the POPE:POPG=3:1 mixture as a function of temperature and compared with the respective heat capacity profiles. It was found that the lifetimes also follow the heat capacity profile. Very similar behavior was found for the sarcoplasmic reticulum calcium channel reconstituted into POPE:POPC mixtures at different ratios \cite{Cannon2003}. Close to the largest heat capacity events of the lipid mixture, the channel displayed maximum activity (highest conductance) and the longest open times.

The behavior described above is expected from the fluctuation dissipation theorem for the lipid transition itself. Both mean conductance and the lifetimes of conduction events accurately reflect the physics of the lipid membrane.  Interestingly, in the case of the KcsA channel the conductance (and thus channel activity) is related to the total amount of protein in the membrane and is still inhibited by the potassium channel blocker tetraethyl ammonium (TEA).  Channel conductance in these systems is apparently a property of the lipid-protein ensemble.  In the following we suggest one possible description for this behavior.

\subsection{A possible catalytic role of membrane proteins}
There is ample evidence in the literature that membrane proteins can influence the thermodynamic properties of lipid membranes. This is true for both  integral and peripheral proteins. Fig. \ref{Figure_14} (center, top) shows the calorimetric profiles of the band 3 protein of erythrocytes, and of cytochrome b$_5$ (Fig. \ref{Figure_14}, center bottom) reconstituted into synthetic lipid membranes \cite{Morrow1986, Freire1983}. While band 3 protein increases the temperature regime of membrane melting in membranes, cytochrome b$_b$ lowers it. Similar observations have been made with other proteins. 
E. \textit{coli} membranes display a lipid transition around 22 $^\circ$C (Fig. \ref{Figure_06} d) while the extracted lipids (in the absence of proteins) display a transition around 12 $^\circ$C.  Since proteins influence melting points, they must also affect lipid membrane fluctuations and the occurrence of lipid channels.   

The influence of an integral protein on membrane melting is partially dictated by the so-called hydrophobic matching \cite{Mouritsen1984}. If the the hydrophobic part of the protein is more extended than the hydrocarbon core of a fluid membrane, it will match better with ordered lipids. As a consequence, lipids tend to be more gel-like at the interface of the protein \cite{Ivanova2003}. If the proteins have short hydrophobic cores (e.g., gramicidin A), it will favor fluid lipids in its vicinity (Fig. \ref{Figure_14}, left). The first class of proteins will shift melting events towards higher temperatures, while the second class will shift it towards lower temperatures. This is shown for the experimental examples Band 3 protein and cytochrome b$_5$ (Fig. \ref{Figure_14}, center). Peripheral proteins can shift transitions, for instance by shielding electrostatic charges on the surface. 

If a protein that matches the gel state of the membrane is located in a fluid membrane, it tends to surround itself by a gel lipid layer. As a consequence, there is a regime of high fluctuations near the protein \cite{Ivanova2003, Seeger2005} meaning that both heat capacity and compressibility can be altered close to a protein. For this reason, the presence of a protein has the potential to locally induce lipid pores in its proximity. Another way of stating this is that proteins can catalyze lipid pores at their outer interface. Fig. \ref{Figure_14}\,(right) show Monte Carlo simulations of such a simulation \cite{Ivanova2003, Seeger2005}. The simulations show a protein (black) in a membrane at temperatures below and above the melting regime. Dark red shades indicate small fluctuations, while bright yellow shades display large fluctuations. It can be seen that a protein that favors the fluid lipid state would tend to create regimes of large fluctuations in its environment at temperatures below the melting temperature (top panels). 

It thus seems likely that proteins can catalyze channel activity without being channels themselves. This is due to the effect of the proteins on the cooperative fluctuations in the lipid membrane. It should generally be possible to estimate this effect from the influence of the protein on melting transitions.

\section{Summary}
The aim of this review has been to characterize lipid ion channels, and to illustrate the similarity of ion conduction events and protein channel activity. We have shown that the appearance of lipid channels is rooted in the fluctuation dissipation theorem. It is thus strictly coupled to the thermodynamics of the membrane and influenced by changes in the thermodynamics variables such as temperature, pressure, voltage, etc. 

Many (but not all) properties of protein channels are practically indistinguishable from lipid channels. These include
\begin{itemize}
  \item Single channel conductances and lifetimes
  \item The current-voltage relations of some proteins such as TRPM8 and the synthetic membrane
  \item The activation of TRP channels by temperature
  \item The conductance and the lifetimes of KcsA and calcium channels embedded in membranes with transitions.
\end{itemize}

However, a few important properties are difficult to reconcile with the pure lipid membrane:
\begin{itemize}
  \item The effect of mutations in the protein.	
  \item The action of strong poisons such as tetrodotoxin or tetra\-ethyl ammonium. Tedrodotoxin in high concentrations (mM regime) only displays a very minor influence on the melting profiles of zwitterionic membranes (unpublished data from Master's thesis of S. B. Madsen, NBI 2012).
  \item The selectivity of the ion conduction, for instance of the potassium channel (about 10000 times higher conductance for potassium over sodium). Lipid channels seem to display a mild selectivity only, following the Hofmeister sequence \cite{Antonov2005}.
\end{itemize} 

Nevertheless, there exist a number of cases, where one can demonstrate clear correlations of protein behavior with the lipid membrane physics, in particular the KcsA channels and calcium channels.

It seems likely that a view will eventually emerge in which the conductance of biomembrane is seen as a feature of a lipid-protein ensemble rather than as a feature of single proteins. This implies a strong coupling to the macroscopic thermodynamics of the biological membrane as a whole.

\vspace{1cm}

\textbf{Acknowledgement:} Thanks to Andrew D. Jackson from the Niels Bohr International Academy for a critical reading of the manuscript. This work was supported by the Villum Foundation (VKR022130).
\vspace{1cm}

\small{

}

\end{document}